\documentclass[sigconf]{acmart}
\AtBeginDocument{%
  }

\copyrightyear{2026}
\acmYear{2026}
\setcopyright{cc}
\setcctype{by}
\acmConference[CSCW Companion '26]{Companion of the Computer-Supported Cooperative Work and Social Computing}{October 10--14, 2026}{Salt Lake City, UT, USA}
\acmBooktitle{Companion of the Computer-Supported Cooperative Work and Social Computing (CSCW Companion '26), October 10--14, 2026, Salt Lake City, UT, USA}
\acmDOI{10.1145/3785651.3831421}
\acmISBN{979-8-4007-2378-0/2026/10}

\usepackage[disable]{todonotes}
\usepackage{url}
\usepackage{xspace}
\usepackage{cleveref}
\usepackage{subcaption}

\begin{document}

\title{\oursys: Bridging Perspectives Across Online Communities with a Dual-Pane Design}

\author{Fei Fang}
\authornote{Johns Hopkins University, Baltimore, Maryland, USA}
\affiliation{}

\author{Reva Hirave}
\authornote{Princeton University, Princeton, New Jersey, USA}
\affiliation{}

\author{William Jurayj}
\authornotemark[1]
\affiliation{}

\author{Yuqi Li}
\authornotemark[1]
\affiliation{\vspace*{-8pt}}

\author{Brian Lu}
\authornotemark[1]
\affiliation{\vspace*{-8pt}}

\author{Tarik Metin}
\authornotemark[1]
\affiliation{\vspace*{-8pt}}

\author{Tsugunobu Miyake}
\authornote{Georgia Institute of Technology, Atlanta, Georgia, USA}
\affiliation{}

\author{Kateryna Morhun}
\authornote{Massachusetts Institute of Technology, Cambridge, Massachusetts, USA}
\affiliation{}

\author{Yash Permalla}
\authornotemark[1]
\affiliation{}

\author{Kenan Rustamov}
\authornotemark[1]
\affiliation{}

\author{Allen Shen}
\authornotemark[1]
\affiliation{}

\author{Haojun Shi}
\authornotemark[1]
\affiliation{}

\author{Prabhav Singh}
\authornotemark[1]
\affiliation{\vspace*{-8pt}}

\author{Xiheng Tom Wang}
\authornotemark[1]
\affiliation{\vspace*{-8pt}}

\author{Kevin Xu}
\authornotemark[1]
\affiliation{}

\author{Qingcheng Zeng}
\authornote{Northwestern University, Evanston, Illinois, USA}
\affiliation{}

\author{Jiayi Zhang}
\authornotemark[1]
\affiliation{}

\author{Daniel Khashabi}
\authornotemark[1]
\affiliation{}

\author{Andrew Perrin}
\authornotemark[1]
\affiliation{}

\author{Tiziano Piccardi}
\authornotemark[1]
\affiliation{\vspace*{2pt}}

\author{Ziang Xiao}
\authornotemark[1]
\affiliation{}

\author{Jason Eisner}
\authornotemark[1]
\email{jason@cs.jhu.edu}
\affiliation{}

\renewcommand{\shortauthors}{Fang et al.}

\begin{abstract}
Social media systems typically display conversations among already familiar contributors, which can be predictable and one-sided.
In civic discourse, this design narrows discussion, reinforces divides, and distorts the perception of public opinion. To encourage cross-community engagement, we present \oursys, an AI-powered bridging system that augments the standard social media feed.  
As the user reads a post, \oursys surfaces diverse relevant posts from other threads in a side pane and highlights the connections.  Users are invited to venture out of their echo chamber, explore a broader range of views and arguments, and ``click across'' to engage with their authors. When they do, \oursys facilitates constructive posting, not only by showcasing relevant past content but also by simulating possible reactions as the user drafts a post.
\end{abstract}

\ccsdesc[500]{Human-centered computing~Collaborative and social computing systems and tools}
\ccsdesc[300]{Human-centered computing~Hypertext / hypermedia}
\ccsdesc[100]{Information systems~Specialized information retrieval}
\ccsdesc[100]{Computing methodologies~Discourse, dialogue and pragmatics}

\keywords{bridging systems; information exploration; applications of LLMs}

\makeatletter
\newcommand*\iftodonotes{\if@todonotes@disabled\expandafter\@secondoftwo\else\expandafter\@firstoftwo\fi}  
\makeatother
\newcommand{\noindentaftertodo}{\iftodonotes{\noindent}{}\ignorespaces}
\newcommand{\fixme}[2][]{{\todo[color=yellow,size=\scriptsize,fancyline,caption={},#1]{#2}}}
\newcommand{\note}[4][]{{\todo[author=#2,color=#3,size=\scriptsize,fancyline,caption={},#1]{#4}}}
\newcommand{\notewho}[3][]{{blue!40}{#3}}
\newcommand{\response}[1]{\vspace{3pt}\hrule\vspace{3pt}\textbf{#1:} }
\newcommand{\jason}[2][]{{green!40}{#2}}
\newcommand{\fei}[2][]{{orange!40}{#2}}
\newcommand{\Fixme}[2][]{\noindentaftertodo}
\newcommand{\Guidance}[2][]{{\todo[inline,color=lightgray,size=\scriptsize,fancyline,caption={},#1]{#2}}}
\newcommand{\Notewho}[3][]{\noindentaftertodo}
\newcommand{\Jason}[2][]{\noindentaftertodo}
\newcommand{\Fei}[2][]{\noindentaftertodo}

\newcommand{\oursys}{\textsc{CrossWeave}\xspace}

\newcommand{\laude}[1]{}
\newcommand{\ournotes}[1]{}%
\newcommand{\defn}[1]{\textbf{#1}}

\maketitle

\section{Introduction}

Social media has been blamed for destabilizing democratic societies.
Large majorities in several countries, and especially in the U.S., believe that social media has made us more divided in our political opinions, less civil in our political discourse, and more vulnerable to misinformation \cite{pew2022socialmedia}.
Yet social media is profitable, popular, and likely here to stay.  So we must meet users and companies where they are.  The challenge is to \emph{mitigate} problems such as echo chambers \cite{cinelli2021echo, sunstein2018republic}, unchallenged misinformation \cite{lazer2018science, vosoughi2018spread},
selective exposure \cite{Bakshy2015exposure, stroud2010polarization}, and perceived polarization \cite{Levendusky2016misperceptions}---and in an appealing way.  

We introduce \oursys, a design that aims to broaden the information diet of social media users, connecting them to a wider range of high-quality ideas that bear on whatever they are currently reading or writing.  It does not intervene in the main social media feed, but rather \emph{augments} it with a curated side pane of social media posts from across various platforms.  Construction of such a side pane is only now becoming possible, using large language models (LLMs) to automatically identify diverse relevant content and highlight connections.  We outline the AI challenges in \cref{sec:ai}, but this paper focuses on the UI concept and its prosocial goals.  

A demo video is at \url{https://youtu.be/QFW9i7KoBhs}.  At time of publication, a live demo is at \url{https://x-weave.org}.

\ournotes{People are stuck in echo chambers and tend to engage with
  like-minded people. The combination of confirmation bias and
  selective exposure leads to a variety of problems like polarization
  (affinity towards like-minded people + diverge from other
  communities), and even if people want to get out of their echo
  chambers, there is no effective system for them to leverage. We are
  designing a \textbf{bridging} system with \textbf{broadening} the
  user’s perspective as our primary goal, to enable users to break out
  of the echo chamber.}

\ournotes{define bridging and broadening respectively, without detailing the design.
E.g. “bridging systems: systems that increase mutual understanding and
trust across divides, creating space for productive conflict,
deliberation, or cooperation.” \url{https://arxiv.org/pdf/2301.09976}}

\ournotes{Not just about polarization, but even within one’s own camp, there’s a narrowness, which motivates the need to broadening}

\laude{
\textbf{We will pioneer} a new social media user interface that uses AI to mitigate polarization, misinformation, and redundancy.  We will augment the standard design with two new features:
\begin{itemize}
\item As users read and write, connect them in real time to relevant \emph{additional} conversations.
\begin{itemize}
    \item[$\Rightarrow$] Goal: Broaden their understanding of the matter they are currently considering, by surfacing diverse high-quality perspectives from \emph{across} the site. 
\end{itemize}
\item As users write, offer them real-time \emph{feedback} on how their post might be received, before they send it.
\begin{itemize}
    \item[$\Rightarrow$] Goal: Improve the quality of argument \emph{within} each conversation.
\end{itemize}
\end{itemize}
}

\laude{We feel it’s our responsibility
  as technologists to fix the mess we’ve created: We propose to
  develop a new social media UI that nudges users to broaden their
  perspectives, by exposing them in the moment to a useful variety of
  arguments from other users.}

\begin{figure*}[!ht]
  \centering
  \noindent
  \begin{minipage}[c]{0.04\textwidth}
    \Large (a)
  \end{minipage}%
  \begin{minipage}[c]{0.76\textwidth}
    \centering
    \frame{\includegraphics[width=\textwidth]{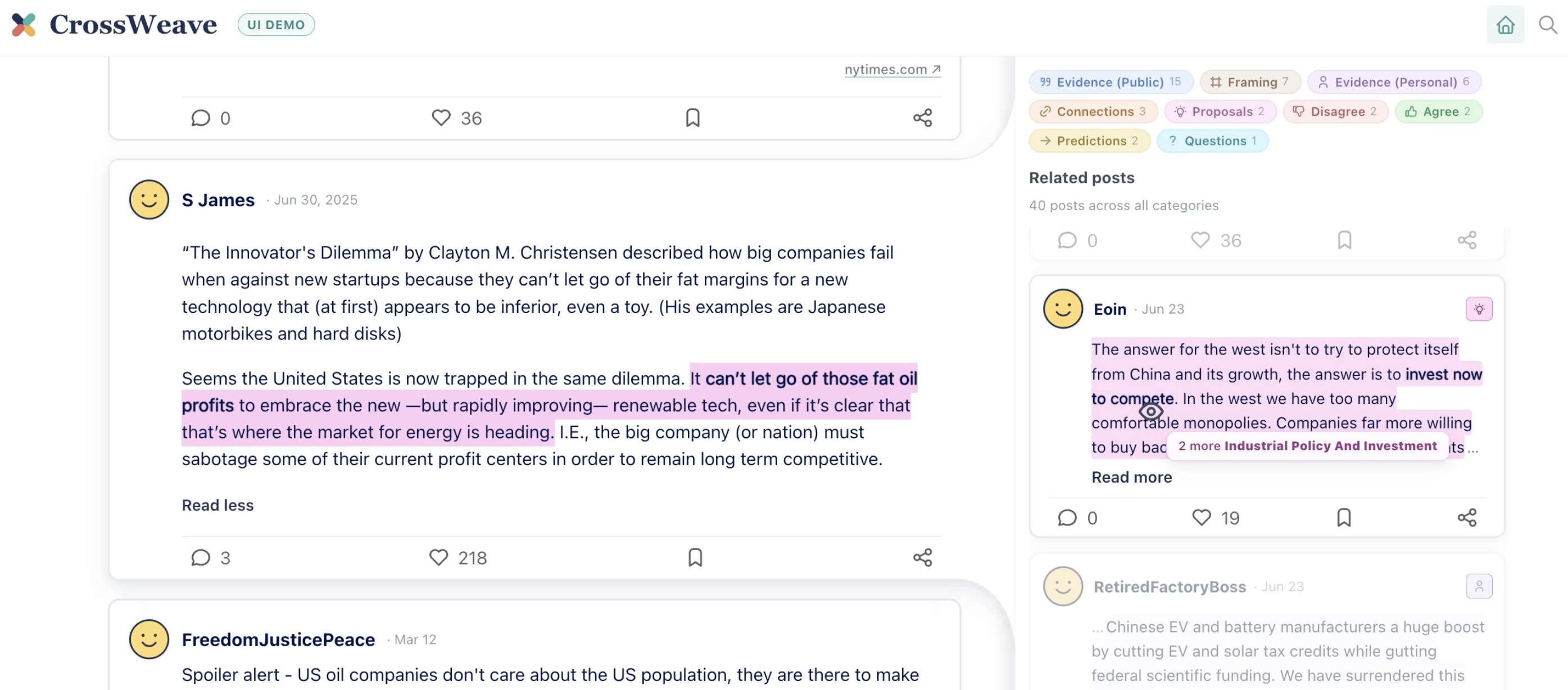}}
    \label{fig:reading}
  \end{minipage}

  \vspace{-2ex}
  \vspace{1ex}

  \noindent
  \begin{minipage}[c]{0.04\textwidth}
    \Large (b)
  \end{minipage}%
  \begin{minipage}[c]{0.76\textwidth}
    \centering
    \frame{\includegraphics[width=\textwidth]{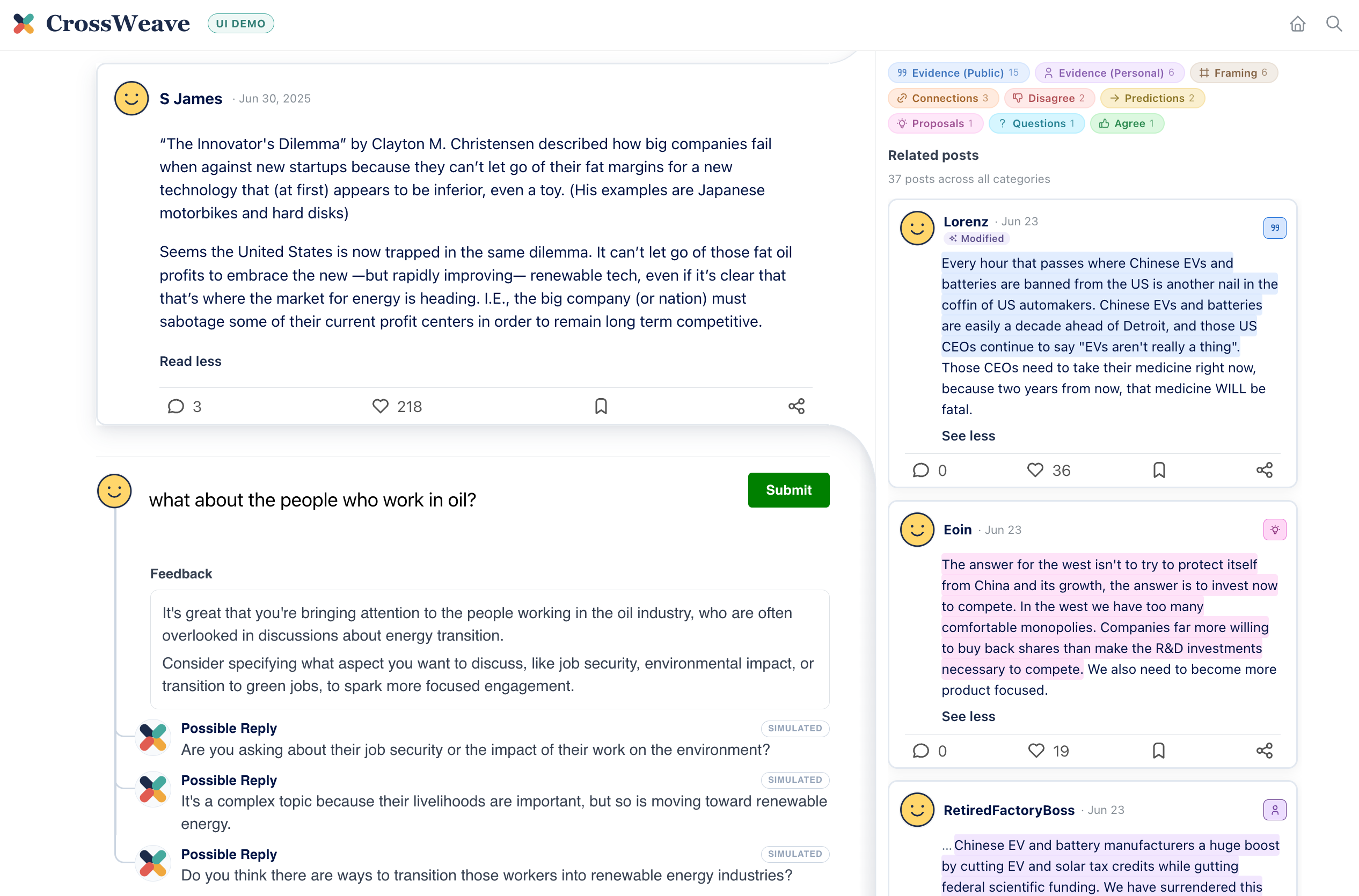}}
    \label{fig:writing}
  \end{minipage}
  \Description{Two screenshots of the \oursys interface, showing ``Reading Mode'' and ``Writing Mode'' for a focused post arguing that the U.S. is trapped in an ``Innovator's Dilemma'' with oil profits versus renewable energy.

(a) Reading Mode: In the main pane, a sentence claiming that the U.S. ``can't let go of those fat oil profits'' is highlighted. In the side pane, a related post by Eoin urging the West to ``invest now'' is shown with a matching highlight and a tooltip reading "2 more Industrial Policy And Investment," indicating that other side-pane posts address the same theme.

(b) Writing Mode: The user is composing a reply in the main pane (``what about the people who work in oil?''). Below the draft, an AI-generated ``Feedback'' box acknowledges the draft for bringing attention to the existing workforce in the energy sector, and suggests bringing up a specific aspect for discussion. Below that, three simulated responses are displayed, e.g., ``Do you think there are ways to transition those workers into renewable energy industries?'' Meanwhile, the side pane shows related posts color-coded by contribution type, including tags for "Evidence (Public)," "Proposals," and "Evidence (Personal)."}

\vspace{-3ex}

  \caption{\oursys's side (right) pane provides scrollable 
  content related to the focused post in the main (left) pane.  (a)~Hovering the mouse over a highlighted span in the side pane reveals the corresponding span in the focused post.  A hover tooltip (``\texttt{2 more \textbf{Industrial Policy And Investment}}'') indicates how many other side posts address the span's topic and invites the user to click to see them. (b)~Without hovering, highlights are visible on all posts in the side pane, color-coded by contribution type (\texttt{Evidence (Public)}, \texttt{Proposals}, etc.).  In this screenshot, the user is drafting a reply: \oursys displays real-time AI-generated feedback and simulates several community reactions, encouraging the user to consider how their post might resonate before submitting.}
  \label{fig:screenshot}
\end{figure*}

\section{Design Principles}

Since the early days of Usenet and Facebook, 
social media and news comment sites have organized discussions into disjoint threads with different readerships.  Arguably, this is the \emph{core} reason for online echo chambers \cite{cinelli2021echo, sunstein2018republic} and for the limited range and quality of many online discussions \cite{Alstyne2005global}. 

\oursys is designed to \textbf{augment} rather than replace this popular traditional social media design.  Sustained conversations within existing social circles are clearly engaging and presumably serve real needs (see \cite{chestertons-fence}).  Users come to social media for many reasons---to maintain offline social connections \cite{Whiting2013why}, to follow and react to news \cite{pewresearch2025social}, to monitor others' reactions \cite{Whiting2013why}, to share their own views and seek validation \cite{naaman2010message}, to attract attention and gain social status \cite{Toubia2013Intrinsic}, to cultivate followers and demolish opponents \cite{Majmundar2018why}, and so on.
They may not be actively seeking information, deep understanding, or new conversation partners \cite{naaman2010message, Toubia2013Intrinsic}. However, by augmenting their display with a side pane (\cref{fig:screenshot}), we hope to make them at least peripherally aware of other views and communities and draw them into exploration and engagement \cite{xkcd-duty-calls-2008,xkcd-bad-opinions-2018}.\footnote{A side pane that surfaces relevant content should have further applications beyond social media.  It might also be useful when reading or writing news stories \cite{yuan2026conewsreader}, academic papers, patents, or legal briefs, and in other discussion forums designed to serve education or democratic deliberation \cite{compdemocracy_polis, pei2025deliberationio,vtaiwan}.}

\oursys is intended as a \defn{conduit} rather than a referee.
All of the content is still written by actual human users: there is no AI voice summarizing comments \cite{wapo-commenting} or ``telling me what to think.''  The side pane showcases a diverse set of views and supports easy exploration, so we hope users will regard it as interesting and empowering rather than biased or coercive.  The side pane is not personalized to a particular user, but only customized to the current context.\looseness=-1

\oursys aims at both informational \defn{broadening} and social \defn{bridging}.  First, it can broaden users' understanding of the current topic by showing them additional perspectives from all angles.
Second, it offers them bridges to new social communities \cite{ovadya2023bridging}: a user can click on any post in the side pane to jump into its original thread and reply if they choose.  

Finally, our design strengthens the \defn{incentive} to \emph{write} useful posts---the sort that will be promoted via the side pane.  A user may put in the effort to rebut a friend's misinformation, knowing that their effort will not be trapped in its original thread, but may be shown again next to similar misinformation when it appears in other threads across the platform (cf. X Community Notes \cite{communitynotes2024introduction}).\footnote{Ideally, the user would be notified and commended when their work is shown elsewhere and draws attention.  This feature is on our to-do list.}

\ournotes{Broadening (engagement with content) — surface broadening content for consumption, balancing quality and diversity in a ranked list.
Highlight: show pairwise highlighted spans as the user navigates down the ranked list.
More specifically: the default is that highlights only show up in the side pane. Upon hovering on a highlight in the side pane, a corresponding highlight shows up in the focused post
The highlighted span in the side pane may also serve as default preview (shown before the user expands to “see more”)
Limitation: the side pane for a given post is the same for all users, though there is room for personalization in the future.}

\ournotes{Bridging (engagement with users).  
Allow users to click into threads to engage with different communities
Facilitate generation of bridging content: incentivize productive conflict/deliberation/cooperation with others when posting/replying, e.g. by simulating replies, surfacing “fodder” for replies (i.e., surfacing existing posts that may provide points of consideration in the drafting process), “kindness reminder” (need to fetch citation)}

\ournotes{Design principles.  
Chesterton’s fence: not disrupting the existing social media experience (including the socializing function with close ones)
How this impacts the implementation of broadening: we keep the main feed (and its ranking algorithm) unchanged, and add a side pane with a ranking algorithm different from that of the main feed to achieve broadening
User agency:  users should feel like they are in control instead of having views shoved into their brains.
How this impacts the implementation of broadening: For ease and agency in navigation, give user the ability to reorder with “See more on topic X” and filter based on contribution type
Maintain perception of light touch and honest brokerage
How this impacts the implementation of bridging: only surface organic content that is minimally modified by AI only for the sake of readability.}

\section{Design Elements}

A \oursys user begins as in other social media interfaces, by scrolling through a list of top-level posts in the \defn{main pane}.\footnote{The main pane may be chronological, an algorithmic feed, or search results.}  They may click on any post to open the conversation that it is part of.  The main pane then changes to show the post with its thread ancestors above it and its replies below it (as in X, Mastodon, or Bluesky).  The user may then post their own reply or continue navigating.

A permanent \defn{side pane} (\cref{fig:screenshot}) displays a ranked list of posts whose contributions relate to the currently \defn{focused post}---that is, the main-pane post that was most recently clicked on or scrolled to.
The user can 
click on any side post to open its thread in the main pane (thus bridging to that related thread).%
\footnote{
Side pane posts have been plucked from their original threads and are sometimes hard to understand out of context (what is ``he,'' ``us,'' or ``this''?).  In these cases, we lightly rewrite the post with AI.  By hovering over the \texttt{Modified} label, the user can see what changes were made (as edit markup) and verify that the meaning was preserved.
}

\subsection{Side Pane Curation}\label{sec:values}

Recall that the entire side pane provides commentary on a single focused post.  Overall, we regard a side pane as good if browsing it would help the reader \emph{reply thoughtfully} to the focused post.\footnote{A possible task for a future user study to evaluate our interface.}  
We select and rank side posts by balancing three criteria: \textbf{relevance}, \textbf{added value}, and \textbf{diversity} (as assessed by AI: see \cref{sec:ai}). 

A side post is \defn{relevant} to the focused post to the extent that it engages with a \emph{specific} claim, stance, or concern raised in the focused post.  That is, the two posts share a \emph{topic}---preferably a fine-grained one.  (Either post may also make additional points.)

A side post \defn{adds value} when it contributes something \emph{beyond} what the focused post already provides.  For example, it may introduce new evidence, offer a fresh argument or counterargument, propose a solution, or suggest a reframing.
In linguistic terms, it supplies a novel \emph{comment} on the existing shared \emph{topic}---much as an ordinary social media reply might \cite{ghosh-etal-2014-analyzing,allen-etal-2014-detecting,benamara2018social}, though the connection may be less apparent in a side post that was not a direct reply.

The side pane as a whole should be \defn{diverse} \cite{carbonell1998mmr}.  As the user scrolls down, successive posts should incrementally add non-redun\-dant value, providing \emph{new} topics, comments, and contribution types.  This directly serves our broadening goals.  A diverse side pane should also be more engaging to users than a list of redundant posts that repeat the same obvious or popular talking points \cite{bar-haim-etal-2020-quantitative}, as we have noticed in the comments sections of many news articles.%
\footnote{By raising the ranking of unusual views, our diversity criterion does risk misleading the user about the prevalence of these views on the platform.  We can mitigate this by ranking posts with popular views higher, other things equal, and also by using other means to indicate prevalence, such as the tooltips in \cref{sec:reading} below.}

\subsection{Enhanced Reading Interface}\label{sec:reading}

How can a user interact with the side pane?  Most simply, they can skim the related posts that are shown, optionally scrolling downward to see more, or clicking to open interesting ones. Beyond this, \oursys includes a suite of affordances to invite substantive engagement with the side pane. In our design, visually salient elements are always invitations to explore.  Accepting an invitation ought to be cheap, rewarding, and non-disorienting.

To draw the user's attention, we reveal \emph{how} each side post $B$ is related to the focused post $A$.  
After all, both posts may also make unrelated points, and $B$ was probably not originally written as a reply to $A$,\footnote{When $B$ is unclear outside its original thread, we use AI to lightly rewrite the side pane's version (e.g., changing ``they'' to ``auto manufacturers'').  We disclose rewriting.} so the $A$--$B$ point of contact is often not obvious.   
To help the user locate it, \oursys highlights one or more text spans in each side post $B$ that justify why $B$ was displayed alongside $A$. These spans were identified by the backend AI.

The highlighted spans are color-coded based on their \emph{contribution types}, which can be one or more of the following: public evidence, personal experiences, connections to other subjects, proposals, predictions, value statements, pointers to further resources, questions, issue framing, humor, as well as agreement and disagreement.

When the user hovers over a highlighted span in $B$,\footnote{Or taps it, on a mobile/touchscreen interface (wide phone or landscape mode).} \oursys temporarily highlights the corresponding span in $A$ (and temporarily dims other highlights in the side pane), as shown in  \cref{fig:screenshot}a.  These two corresponding highlighted spans generally contribute different \emph{comments} on the same topic, which \oursys calls out by temporarily bolding phrases within them.

Conversely, the user can also hover over some interesting or questionable part of $A$.  Although $A$ does not show any highlights by default (to minimize reading distraction), hovering over $A$ will reveal its bold phrases.  Hovering over a \emph{specific} bold phrase will show its topic as a tooltip, and will temporarily dim all off-topic highlights in the various $B$ posts.  \emph{Clicking} on the bold phrase filters the side pane to just the $B$ posts with on-topic highlights.%
\footnote{If a bold phrase has multiple topics, then clicking rotates through them and also the tooltip becomes a clickable selection menu.}

The posts in the side pane can also be filtered by contribution type, via clickable icons on each post
and at the top of the pane.  The side pane normally shows a diverse mix of contribution types.

Finally, we let the user override the AI's ranking in the side pane.
While the initial ranking favors diversity (breadth-first), sometimes the user may be curious to dive (depth-first) into a particular topic raised in a side post $B$: ``Are others making this point and how did they phrase it?''  Thus, when the user hovers over a highlighted span in $B$, a tooltip like ``\texttt{8 more \textbf{AI Hype}}'' signals the prevalence of posts making this point and hints that the user can click to see more of them.  Clicking will raise the ranking of the next 3 ``AI Hype'' side posts so that they fall immediately below $B$.  The user can scroll through this topical block of posts, then click a link at the bottom to expand the block further with 3 more, or else just scroll past the current end of the block to resume reading other posts.  All posts within the block highlight their ``AI hype'' spans, with the other highlights dimmed.\footnote{The side pane allows at most one topic block at a time.  Canceling the current topic block or switching to a new one removes the current block formatting, but leaves the posts in their position so that the user does not have to encounter them again when scrolling downward.}

While these features are primarily designed to facilitate navigation of the side pane, we also support them for the focused post's direct replies in the main pane. This should help when a post attracts a large number of replies (e.g., a breaking news article from a trusted source). Rather than reading replies in chronological or popularity order, the \oursys user can efficiently broaden their view by surveying the diverse perspectives scattered across hundreds of responses.

\laude{When the user hovers over any related post, we will textually highlight the relevant topic in the focused post, and simultaneously highlight the related post's contributions on this topic.
The goal is to orient the user quickly.  (Posts may be multifaceted, and the related post was not originally written as a reply to the focused post, so the point of contact may not be immediately obvious.)
\item[Fine-grained relatedness]
Conversely, hovering over different parts of the \emph{focused} post will highlight different topics.  Clicking on one of them will filter the side pane to show only related posts that comment on that topic.}

\ournotes{Broadening (engagement with content) — surface broadening content for consumption, balancing quality and diversity in a ranked list.
Highlight: show pairwise highlighted spans as the user navigates down the ranked list.
More specifically: the default is that highlights only show up in the side pane. Upon hovering on a highlight in the side pane, a corresponding highlight shows up in the focused post
The highlighted span in the side pane may also serve as default preview (shown before user expands to “see more”)
Limitation: the side pane for a given post is the same for all users, though there is room for personalization in the future.}

\laude{We will provide several AI-powered features for easily
  navigating the side pane.  As a result (we hope), users will feel
  empowered to explore different views, rather than coercively
  steered.}

\laude{
  \begin{description}
\item[Ranking] 
\item[Prevalence] Highly ranked posts may express either common or rare views.  To give users a realistic sense of the political landscape, each related post visually indicates the \defn{prevalence} on our site of the contribution it makes.  We have yet to design this indicator, but it should be stronger for a point that is often made or often upvoted.
\item[Exploration] 
The user can also filter the side pane by contribution type, by clicking icons at the top of the side pane.  
We have identified several types of contributions: public evidence, personal experiences, arguments, pointers to further resources, proposals, predictions, value statements, connections to other ideas, questions, issue framing, humor.
\item[More like this] The user can override the diversity ranking criterion to dive into a particular topic.
If the user discovers an interesting topic in a related post while scrolling down the side pane, they can click the topic's prevalence indicator to see a few more related posts commenting on the same topic. That is, these posts' ranking will be raised: they now appear immediately under the clicked post, rather than farther down in the side pane.  The user can read them and scroll past them, or click again to bring up additional related posts.
\end{description}
}

\subsection{Enhanced Posting Interface}

\oursys should also improve average content quality on the platform.  As the user drafts a reply to the focused post, the side pane of related posts 
provides food for thought, as  \cref{sec:values} noted.
Some related posts may provide counterarguments that the user may care to address; others may offer supporting evidence or interesting angles that the user could incorporate.
The user can quote a related post by dragging it into their own draft.

We also surface real-time feedback as the user writes (\cref{fig:screenshot}b). It is generated by an AI persona \cite{kong-etal-2024-better}--a ``wise and patient teacher and disussion moderator''---who both acknowledges the contributions of the current draft and provides suggestions to further strengthen its articulation.  \oursys also previews how the current draft might resonate with various communities, by displaying three AI-simulated replies to the draft---each adopting a distinct persona---and  disabling submission for several seconds. With this preview, the user is encouraged to assess whether they would be satisfied with how the post might be perceived, and revise accordingly if not.\looseness=-1

\laude{Beyond raising the quality of what users get to \emph{read}, a second goal is to raise the quality of the posts and replies that they \emph{write}.}

\laude{So we will not try to steer the user toward some academic ideal of polite, informed, balanced discourse.  Users would surely resent and resist such heavy-handed attempts.  In fact, opinionated or intemperate posts that clearly make a single point can be valuable.}

\laude{Rather, we will try to support the user's own communicative goals.  Our role is that of a blind-spot mirror, installed to provide a new angle and show the driver what they might otherwise miss. We provide three kinds of \emph{private} feedback as the user is writing:
\begin{description}
\item[Related posts] We will again use the side pane to surface related posts---that is, existing posts related to the user's draft.  The user may glance at these and can explore them further if they seem useful.  Some related posts may object to the very arguments and evidence that the user is presenting.  Others may provide supporting arguments and evidence, or useful connections.  The user can quote and link to a related post by dragging it into their own post.
\item[Synthetic replies] Any future replies to the user's post will appear indented below it in the conversation.  We will preview how readers might react by showing a few brief AI-generated replies in this space (clearly marked as such).  This can alert the writer when others might be confused by their post, or when others might call them out for misinformation, omissions, or blindness to relevant concerns.  AI-generated replies can also provide positive reinforcement and useful material.  They are much like related posts, but are directly targeted to the user's draft in its conversational context.
\item[Advice] We may also provide a few sentences of advice (including praise) generated by a helpful AI persona that is looking over the user's shoulder.  As the user updates their draft, we will revise this space to react in real time to their changes, as in a dialogue. 
\end{description}}

\section{Data and Goals for the Live Demo}\label{sec:data}

Our demonstration focuses on conversations about U.S. news (rather than about books or movies, say).  Our demo system is currently populated with over 300 reader comments harvested from Reddit, Fox News, and The New York Times.%

We invite users of the demo to consider whether \oursys could improve civic discourse on social media:
\begin{enumerate}
\item Does the system feel trustworthy and usable?
\item When you are reading a post, does the side pane add value by genuinely broadening your perspective?
\item Would the side pane and the posting feedback help you take more informed and defensible positions on contested political issues, given your \emph{own} values?
\item Would the system help you understand your political \emph{opponents'} positions, arguments, and values well enough that you can have more constructive dialogue with them?  Would it draw you into such dialogue?
\end{enumerate}

\laude{What makes a post ``related''?  
It can \emph{serve as a kind of reply} to the focused post.  More precisely, 
it must address a topic or claim that was explicitly or implicitly raised in the focused post.  Furthermore, its contributions on that topic should be new to the discussion.}

\laude{We have a cascade of goals, each resting on the previous one:
\begin{enumerate}
\item 
Users should trust our system and like using it.  They should feel that it gives them easy access to interesting content. 
\item Users should develop more informed and defensible positions on contested political issues, given their \emph{own} values.
\item Users should come to understand their political \emph{opponents'} information, position, and values well enough that they can have more constructive dialogue with them. \ournotes{move "constructive dialogue" into next point?}
\item \ournotes{improved posting on the site itself, not just at Thanksgiving}
\end{enumerate}}

\laude{Our UI design aims to connect each user to diverse and
  interesting \emph{additional} conversations that are \emph{relevant}
  to whatever they are reading right now.  We aim to broaden the
  user’s perspective on what they’re reading.  There’s no reason to
  limit them to the posts that happen to be in the same thread.}

\laude{[not for this doc] \textbf{We will evaluate} the impact of these changes on users'
  \emph{political} understanding and attitudes.  We will also study
  whether users accept this interface as a fair source of political
  news, use the new features as intended, and actually prefer the
  experience.}

\laude{We will use LLMs and other machine learning to help curate the
  social media display.  When the user is focused on a post or reply
  in the traditional feed, we will surface high-quality \emph{related}
  posts in the margin, \emph{grouped} and \emph{highlighted} to
  connect them to the focused text.  Users can glance at the related
  posts, browse through them, and click on them to explore them in
  their original context.}

\laude{These related posts will be chosen by AI methods that we will
  develop.  The related posts are drawn from users across the entire
  site, so they break through the filter bubble and provide other
  relevant perspectives---supporting and opposing evidence,
  proposals, connections to other ideas, counterarguments, personal
  stories, opinions, clarifications, etc.}

\laude{We won’t put a thumb on the
  scale: real users in their own words.}

\laude{Users now have more incentive to write high-quality posts
  (e.g., correcting misinformation) because those posts will be widely
  featured wherever relevant.  Our UI will also help users construct
  better posts by providing real-time feedback (related posts, advice,
  and simulated replies) as they write their drafts.}

\laude{AI curation is more scalable than manual comment moderation, and can be much more ambitious, as our design shows. Even basic manual moderation is expensive and difficult.  Many news sites have abandoned moderation or disabled comments altogether \cite{journalmedia2040034}.}

\ournotes{we could frame the above as a problem of steering engagement. There are two types of engagement:
Engagement with content → broadening
Engagement with user who posts the content → bridging
How do we scaffold both types of engagement?
Sourcing content for consumption from across communities presents the opportunity to engage with those from other communities. 
There’s a set of features oriented towards “better writing” that supports/facilitates the engagement once a user has decided to engage.}

\section{AI Challenges}\label{sec:ai}

Our demo uses precomputed side pane data for each possible focused post. We first use an LLM (currently GPT-5.4) to describe and classify the contributions of each post, to identify a highlightable span for each contribution, and to assess the relevance and added value. Then, from these judgments, a cheaper secondary pipeline (GPT-5.4 and GPT-4.1-mini) induces a small set of subtopic labels to enable the depth-first drilldown. To finalize the ranking, we first use a deterministic, interpretable function to combine relevance and added value into a single score, and then rank greedily via maximal marginal relevance \cite{carbonell1998mmr}, penalizing redundancy in subtopic and contribution type.

While this implementation illustrates the UI vision and confirms that AI can indeed be used to identify and highlight relevant content, it is quadratically expensive and runs offline.  Handling live traffic would require \oursys to construct each side pane efficiently on demand (at least when no recent version is cached).  We have been working on this problem for some time, in parallel to developing the UI design. Challenges include scalability and latency.  A reasonable workflow would identify high-quality reusable posts as they arrive, preprocess them by extracting their substantive key points (see the contribution types in \cref{sec:reading}), semantically index these key points for rapid retrieval from a precomputed topic graph or vector database, and later rerank retrieved posts on the fly using a fast scoring function that considers diversity \cite{carbonell1998mmr}.

Of course, rapidly assembling a diverse list of relevant documents is the central task of information retrieval.  IR and NLP researchers have recently been incorporating LLMs into the indexing process \cite[e.g.,][]{gao-etal-2023-precise}.
Still, our criteria in \cref{sec:values} may be particularly hard to index for.\footnote{For example, a real reply might challenge a \defn{virtual standpoint} that was only quietly assumed by the focused post \cite{vaneemeren2002pragmatic}.  Finding non-replies that can play this role seems especially difficult using only vector-based retrieval and reranking, even if a powerful LLM could read them in the context of the focused post and recognize their relevance.}  They seem to require closer reading, deeper reasoning, and more structured human evaluation than traditional IR settings.  We find that it is challenging even for a human to assess numerically how relevant a post is, how valuable its contributions are, and how distinct they are from the contributions of other posts.  Furthermore, such assessments are value-laden, subjective, and likely gameable.  Yet we must take care to make them fairly, so as not to inadvertently suppress some viewpoints or lose user trust on suspicion of doing so.  
In short, we regard the retrieval of broadening content as a challenging new task for NLP.

\laude{Civics and rhetoric: Defining our criteria for curating the user’s display.}

\laude{Workflow engineering: Operationalizing these criteria by using LLMs, human annotators, and user feedback, including ways to elicit annotations and handle human disagreements and wayward LLMs.}

\laude{Machine learning: Making the system acceptably fast.  It’s not feasible to call humans or LLMs to assess each possible pair of related posts at runtime.  Rather, we need to distill our criteria into a more efficient system that uses precomputed vectors for dense retrieval and reranking.}

\laude{The AI here is challenging, even though LLMs are skilled at
  parsing social media.  It’s hard even for a human to decide which
  posts would add the most value to the current context.  A single
  post can be complex, heterogeneous, or confusing even in its
  original context.  Most posts will be off-topic; those that do share
  a topic with the current post may not respond to its exact points;
  and even the responsive ones may not add real value.  Furthermore,
  we can’t just consider the posts one at a time, because we want to
  expose a variety of arguments on different sides while also
  accurately conveying the prevalence of different views.}

\laude{Our technical approach includes several elements:
\begin{description}
\item[Large language models (LLMs)] make this project possible.  With careful prompting, LLMs can do a sensitive close reading of each post within its original context \cite{lee2025pragmatic, yi2025irony}.  They also provide good interactive feedback on draft posts \cite{argyle2023chat, yeo2024help}.
\item[Information retrieval (IR).]  Given the close readings, we will train neural networks to rapidly find posts that will broaden the user's understanding along varied dimensions. 
\item[User interface design (HCI).]
Given the currently focused post, we will present broadening material off to the side, aiming to make it attractive and convenient to browse. We will highlight its relevance to what the user is currently reading, and organize it to encourage curiosity-driven exploration of high-quality relevant content.
\item[Annotation.] Selecting and organizing the posts requires many steps.  To train our machine learning systems to carry out these steps wisely, we will develop active learning workflows that obtain and integrate supervision of various kinds from human annotators, end users, and LLMs. 
\end{description}
}

\ournotes{\section{Related Work}}

\ournotes{mention "latitude of acceptance"}

\laude{The closest precedent we know of is X Community Notes \cite{communitynotes2024introduction}, where a trusted rebuttal is displayed alongside all posts sharing the same image or link.  However, X uses this mechanism only for rebuttals to false or misleading posts, and limits it to rebuttals that have been written and vetted by anonymous volunteer moderators (under 10\% of these submissions actually get approved, and it takes 14 days on average).  Thus, most posts on X are still presented without any broadening context \cite{braga2025community}.  We address this by routing \emph{organic} rebuttals and comments to a suitably wide audience.}

\laude{Other recent work on designing ``bridging systems'' for social media \cite{ovadya2023bridging} has focused on the traditional feed ranking algorithm.  It aims to promote individual posts that have bipartisan appeal, rather than polarizing appeal \cite{tornberg2023simulatingsocialmedia}.  
We are likewise interested in finding good posts to promote.  However, our approach is orthogonal:
\begin{itemize}
\item We optimize the \emph{list} of posts in the side pane so that it exposes the user to a range of views and arguments.  It is not necessary for each \emph{individual} post in the list to have bipartisan appeal.\footnote{\label{fn:quality}\label{fn:personalization}That said, bipartisan appeal is not irrelevant. Our ranked list considers quality as well as diversity.  Among other quality factors, we may rank posts higher if we think they are accurate, valid, constructive, and 
able to resonate with the \emph{current user} \cite{Sherif1961social} rather than risking a ``backfire effect'' \cite{bail2018exposure,swire2022backfire} (\S\ref{sec:scaling}).}  
\item Modern LLMs can do close reading and identify detailed arguments.  As a result, we can show three distinct reasons to support a proposal, interleaved with three distinct reasons to oppose it. Prior work was limited to much coarser bridging constructs, such as assessing whether a post is constructive by analyzing user reactions or surface language \cite{ovadya2023bridging}.\looseness=-1
\item We look at whatever the user happens to be reading right now---perhaps a post written or recommended by a friend---and provide broadening context for it: ``What should I think about what I am now reading?''  This is quite distinct from reranking the main feed: ``I'm bored, what should I jump to next?''  While both questions are important, we believe that providing broadening context in a side pane is an important new idea with wide applicability.
\item Our interface can promote reply posts, not just top-level posts.  Replies are a rich source of broadening content---e.g., counterarguments and anecdotes---since their original intent was to respond to something.
The user can click to view these in their original conversational context---breaking out the filter bubble and bridging to new perspectives.
\end{itemize}}

\section{Conclusion}

\ournotes{Section 4. Conclusion: implications, expected effect, what studies does this open? Relevance to CSCW}

Social media is a consequential user interface.  \oursys is our effort to reimagine it with AI, not by adding a chat pane but by designing new helpful and prosocial features that are quietly enabled by LLMs.  \oursys attempts to connect isolated social media threads to the broader landscape of viewpoints, arguments, and users, in the hope of giving users a more mature understanding of this landscape and enabling them to participate more constructively through their own posts.

\laude{Traditional social media design (Facebook, Threads, Instagram, Usenet, Reddit, X, Nextdoor) displays each post only within its original conversation.  Although we will preserve these threaded conversations, we will situate them in a broader landscape of viewpoints and supporting information, populated by liberating other posts from \emph{their} conversations.}

\laude{In doing so, we hope to showcase better discourse---and thus incentivize it.  On Facebook, there is little point in spending an hour carefully rebutting your friend's argument, because only your friend's followers will see your reply.  On our platform, your rebuttal can spread further if the AI recognizes it as valuable to also display alongside \emph{other} posts similar to your friend's.}

\begin{acks}

We thank Sanjeev Khudanpur and the 10th Frederick Jelinek Memorial Summer Workshop on Speech and Language Technology (JSALT 2024) for hosting and funding the 6-week residential hackathon in which we developed a previous version of this interface.  Our project has subsequently been funded by a 2025 Johns Hopkins University Discovery Award, ``Using AI to Make Social Media More Prosocial.'' Our gratitude also goes to Evan Prodromou and Andy Piper from Mastodon and to the authors of many open-source tools.  

As \cref{sec:data} mentions, our demo makes use of harvested public comments.  We thank the authors of those comments and reproduce them in our demo under fair use, with their username shown.  The official ACM version of this paper has supplementary material that credits the authors of example comments from the official paper and video.
\end{acks}

\bibliographystyle{ACM-Reference-Format}
\bibliography{crossweave}

\end{document}